# Cavity-enhanced narrow bandwidth photon pairs at telecom wavelength generation with a triple resonances configuration

Zhi-Yuan Zhou[1,†], Dong-Sheng Ding[1], Yan Li[1], Fu-Yuan Wang[1], Bao-Sen Shi[1,*] and Guang-Can Guo[1]

[1]*Key Laboratory of Quantum Information, University of Science and Technology of China, Hefei 230026, China*

E-mail: * drshi@ustc.edu.cn

† zyzhouphy@mail.ustc.edu.cn

**Abstract.** In this paper, we report on the preparation of a narrow bandwidth photon source at the telecom wavelength using spontaneously parametric down-conversion in a type II PPKTP crystal inside an optical cavity. Simultaneous resonances of the pump and the two down-converted fields are achieved by the special design of the cavity. This triple resonances optical parametric oscillator operates far below threshold, generating photon pairs at the wavelength of 1560 *nm* with a bandwidth of 8 *MHz*. A coherence time of 27.7 *ns* is estimate by a time correlation measurement, and the estimated production rate of the photon pairs is 134 $s^{-1}MHz^{-1}mW^{-1}$. As a photon pair in the telecom regime is suitable for long-distance transmission, this source is desirable for the demands of quantum communication.

## 1. Introduction

Long distance quantum communication needs to transmit entangled photons between remote parties. The transmission distance is limited by photon absorption in the optical fiber, which increases exponentially with the distance. To overcome this limitation, a few quantum repeater protocols have been proposed [1, 2]. By using quantum repeaters, the entire communication distance can be divided into smaller segments across which two quantum memories [3] can entangle effectively. By swapping operations on the intermediate nodes, the entanglement can transmit along the whole network.

To realize the above schemes, a narrow bandwidth photon pair source is required. First of all, because the natural spectral linewidth of the atom is at the order of MHz, the photon with narrow band can couple to atomic-based quantum memories [4, 5] efficiently. Moreover, the photon with the narrow band have a long coherence time, the tolerance of the fiber length fluctuations and chromatic dispersion could be increased if it transmits in the fiber as the information carrier.

Spontaneously parametric down-conversion (SPDC) [6] in nonlinear crystals is demonstrated more practically with respect to other techniques, such as Raman scattering or spontaneously four-wave mixing in atomic ensembles [7-11], and bi-exiton cascade in quantum dot [12, 13]. However, due to the broad spectra of SPDC source (of the order of THz), the bandwidth needs to be reduced significantly. One common used technique is to place a nonlinear crystal inside an optical cavity [14-25]. In the process of intra-cavity SPDC, the bandwidth of emitted photon is decided by the cavity, and the SPDC process is enhanced compared with the single-pass scheme at the same time [14, 15].

The initial reported OPO sources produced photon pairs in the visible regime between 700 *nm* to 900 *nm*. In [21], a narrow band single-mode polarization-entangled photon source with a bandwidth of 9.6 MHz was reported, and the first long-term stable photon pairs OPO, with a bandwidth of 3 MHz at 894.3 nm was made in [23, 24 ]. Only recently, OPO source in the telecom regime based on PPLN waveguide was reported [26], where a narrow band photon source with bandwidth of 117 MHz at 1560 *nm* was prepared. However, the bandwidth of the photon pairs in the telecom regime need further reduce to meet the requirement of long-distance quantum communication.

In this paper, a narrow bandwidth photon source in the telecom regime based on a type II PPKTP nonlinear crystal is prepared. Our system is the first realization of narrow band photon source in the telecom regime using bulk OPO. By locking the cavity to the pump laser, and by finely tuning the temperature of the crystals, we realize the simultaneous resonances of pump and the two down-converted fields. The bandwidth of the photon generated is 8MHz and the estimated spectral brightness of our OPO source is $134 s^{-1} MHz^{-1} mW^{-1}$. In the first part of this paper, we show the properties of PPKTP crystal by performing a single-pass SPDC experiment. Then we put the crystal into an optical cavity. The parameters of the cavity, the tuning behavior and stability of the triple resonance cavity are described in details. Finally, time correlations of the photon pairs in the multi-mode case and single-mode case are measured respectively. The main results are discussed in the conclusion section.

## 2. Details of the source

### 2.1 The crystal

The type II PPKTP crystal we used is bought from Raicol Crystals, with a dimension of $1\times2\times10\ mm^3$, the crystal is periodically poled with a periodicity of 46.2 *μm* to get quasi phase match (QPM) for SPDC at a pump wavelength of 780 nm and signal and idler wavelengths of 1560 *nm*. Both end faces of the crystal are anti-reflect coated at wavelengths of 780 *nm* and 1560 *nm*. The crystal is x-cut for pump beam propagation along the x axis of the crystal. The crystal is designed for type-II phase match of $n_y(\omega_p) \to n_y(\omega_s) + n_z(\omega_i)$, where $n_{y,z}$ is refractive index along the y and z axis of the crystal respectively, $\omega_{p,s,i}$ is the frequency of the pump (*p*), signal(*s*) and idler (*i*) field respectively. The photon pairs generated are orthogonally polarized. The temperature of degenerate down-conversion is determined by second harmonic generation (SHG) using a diode laser (DL prodesign, Toptica). The wavelength of this laser can be continuously tuned form 1520 *nm* to 1595 *nm*, with an output power of 30 *mW* at 1560 *nm*. The temperature dependence of the SHG power is showed in Figure 1, where the green stars are experimental measured data, and the red line is fit using $\sin c^2$ function. From the measured data, we could see that the optimal phase match temperature is around 22.0 ℃, and the temperature bandwidth of SHG is about 80℃.

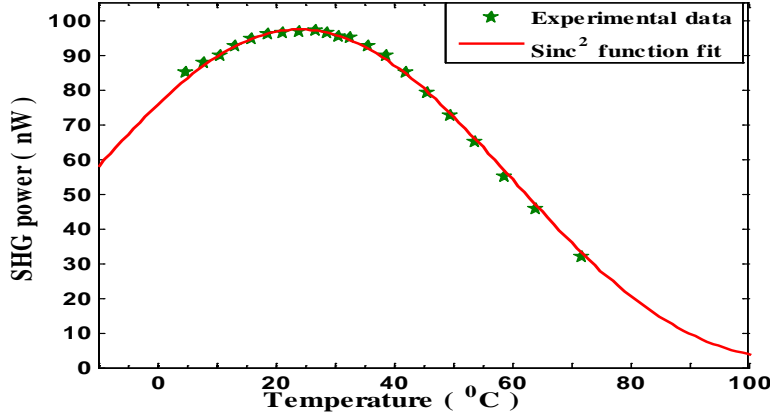

Figure 1. SHG power as the function of temperature. The green stars are experimental data, red line is fit using $\sin c^2$ function

## 2.2 Single-pass experiment

In order to investigate the qualities of this crystal and the parameters of our photon counting system, we perform a single-pass SPDC experiment firstly. The schematic setup is depicted in Figure 2. The pump light at wavelength of 780.007 *nm* from a CW (single-mode, highly stabilized with less than 75 kHz line width) Ti: Sapphire laser (Coherent MBR 110) is focused into the crystal using a lens L1 with 150 *mm* focal length. The polarization of the pump light could be changed by a half wave plate placed in front of the crystal, it is removed by a band-pass filter with a bandwidth of 10 *nm* after crystal. The down-converted signal and idler photons are collected using a lens L2 with 50 *mm* focal length. They are departed using a polarization beam splitter firstly, then collected using two single-mode fibers (SMFs), and sent to two InGaAs Avalanche Photodiodes ( APDs, Qasky) .

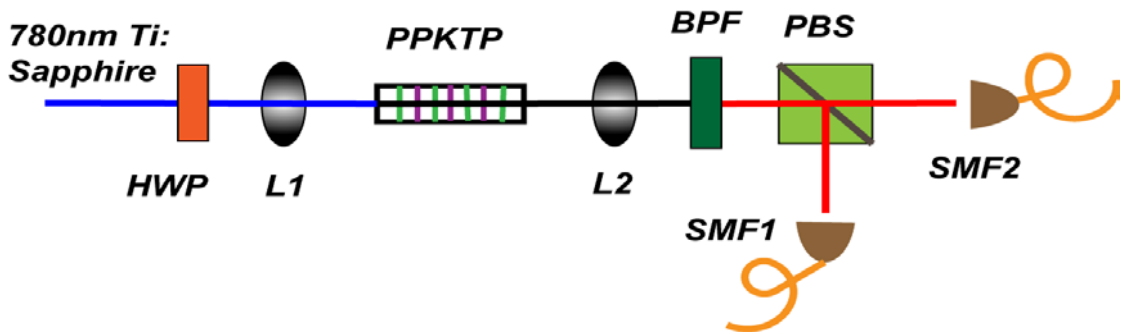

**Figure 2.** Experimental setup of the single-pass SPDC experiment. HWP: half wave plate; L1, L2: lens; BPF: band pass filter; PBS: polarization beam splitter; SMF1, SMF2: single mode fiber.

The two APDs are all used in the gated mode with a detection efficiency of 8%. The first APD is running in the 10MHz internal trigger mode with 1$\mu s$ dead time, 5 $ns$ detection window and $6\times10^{-6}$ $ns^{-1}$ dark count rate. The second APD is triggered by the output signal of the first APD with a detection window of 2.5 $ns$. The photon sent to the second APD is optically delayed by about 1 $\mu s$ using a 200-m long single-mode fiber. The output signal of the first APD is electrically delayed by a delay generator (DG535, Stanford). The output signal of the second APD is sent to a counter to record the coincidence. The temperature of the crystal is controlled by a homemade temperature controller with a stability of ±2$mK$.

The polarization of the pump beam is rotated by a half wave plate to make it polarized along the y axis of the crystal (horizontal direction for our setup). The temperature of the crystal is controlled at 22℃ to get degenerate SPDC outputs. We measure the single counts and coincidence counts as a function of pump power, the dark count and accidental coincidence are subtracted. The results are showed in Figure 3, it shows that the single counts and coincidences are linearly proportional to the pump power. Consider the collecting efficiency (30% to 40%) of the two SMFs and the transmission of the band-pass filter (70%), we estimate the pair production rate to be about $2.66\times10^4$ pairs(s·$mW$)$^{-1}$ using the formula

$$R_{estimate} = R_{detected} / (d\alpha_1\alpha_2 t^2\eta^2).$$

Where $R_{estimate}$ is the estimated pair production rate and $R_{detected}$ is the detected pair rate, $\alpha_{1,2}$ is the fiber collection efficiency of signal and idler photons respectively, $t$ is the transmittance of the band-pass filter, $\eta$ is the detection efficiency of APD and $d$ is detection duty cycle of each trigger period, which is $d = 2.5\%$ for our detection system.

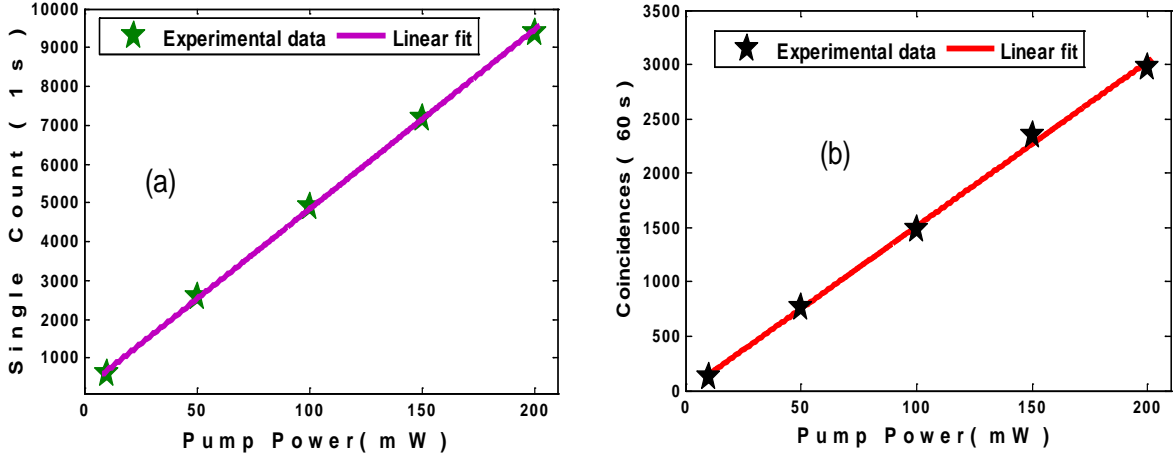

**Figure 3.** Single count per second (a) and coincidence count per 60 seconds as a function of pump power. The dark counts and accidental coincidences are subtracted.

## 2.3 The parameters of cavity

The OPO cavity consists of two concave mirrors and two piece of crystals: a 10-mm long PPPKT and an additional 10-mm long KTP crystal with its axis rotated by $90^0$ for compensating the free spectral differences of the signal and idler photon inside the cavity. Both end faces of the two crystals are AR coated at 780 $nm$ and 1.56 $\mu m$. The radius of both concave mirrors is 78.9 *mm*. The input mirror is high-reflectance coated (R>99.8) at 1560 *nm*, and has a transmittance of 4% at 780 *nm*. The output mirror is high-reflectance coated at 780 *nm* (R=99.86%) and has a transmittance of 3.3% at 1560 *nm*. The length of the cavity is 122 *mm*. We measure the free spectral range (FSR) and bandwidth of cavity for the pump beam at 780 *nm* and down-converted beam at 1560 *nm* by scanning the cavity. The transmission spectrum of them are detected by fast photodiodes and recorded by an oscilloscope. The acquisition wave-forms are showed in Figure 4. The measured FSR and bandwidth for the pump beam (down-converted beam) are 949 MHz (952 MHz) and 10 MHz (8 MHz), respectively. The two crystals are mounted on two stages, each one can be aligned by two-axis tilting. The beam waist is about 120 *μm* in the cavity. The temperatures of them are controlled by two temperature control units separately. The fluctuations of the crystal temperature are limited to ±2*mK*. The output mirror is attached to a piezoelectric transducer (PZT), the length of the cavity is actively stabilized to the pump beam via Pound-Drever-Hall method (PDH)[27].

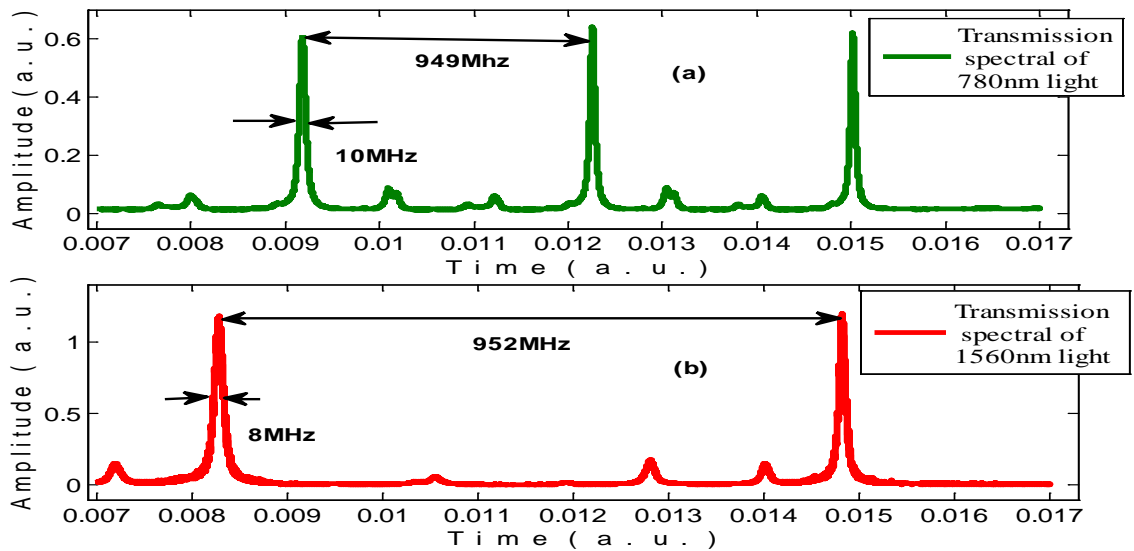

**Figure 4.** Transmission spectrum of (a) the pump at 780 nm and (b) down-converted beam at 1560 nm.

## 2.4 Experimental setup

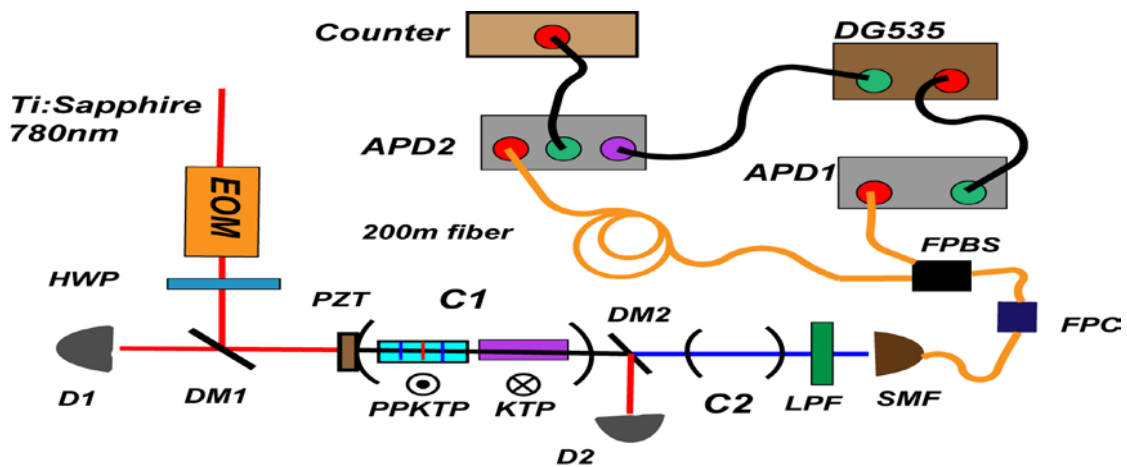

**Figure 5.** Experimental setup of the OPO system. C1: main cavity; C2: filter cavity; EOM: electrical optical modulator; HWP: half wave plate; DM1, DM2: dichromatic mirror; D1, D2: 1560nm, 780nm photodiode; PZT: piezoelectric transducer; LPF: long pass filter; FPBS: fiber polarization beam splitter; FPC: fiber polarization controller; SMF: single mode fiber; APD1, APD2: Avalanche Photodiodes; DG535: delay generator.

The CW single-mode frequency-locked pump beam at 780 *nm* is modulated by an electrical optical modulator (EOM) to generate sidebands for generating an error signal to lock the cavity, then it is coupled into the main cavity (C1). Photodiode D1 is used to detect the transmission spectrum of 1560 *nm* light. The leaked pump beam of the cavity is reflected by a dichromatic mirror (DM2) and detected by a photodiode (D2). The signal of the D2 is sent to a servo control unit to actively stabilize the length of C1 using PDH method. The leakage of the pump beam from the output mirror is removed by DM2 and a long-pass filter (LPF) (cut-off wavelength is 1000 *nm*). The signal and idler photons are filtered by a filter cavity C2 (30 GHz FSR, finesse of 100), the filtered photons are coupled into a single-mode fiber (SMF), and then separated using a fiber polarization beam splitter. After that, these photons are sent to a coincidence record system, which is the same as that used in the single-pass case.

## 3. Experimental results

### 3.1 Multi-mode output

The results without filter cavity C2 will show firstly. We measured the temporal intensity cross-correlations of the signal and idler photons. The cross-correlation function for a type-II double resonance SPDC is given by [23, 28]

$$G_{s,i}^{(2)}(\tau) = \langle \Psi | E_i^{(-)}(t) E_s^{(-)}(t+\tau) E_s^{(+)}(t+\tau) E_i^{(+)}(t) | \Psi \rangle$$

$$\propto \left| \sum_{m_S, m_I = -\infty}^{\infty} \frac{\sqrt{\gamma_S \gamma_I \omega_S \omega_I}}{\Gamma_S + \Gamma_I} \begin{cases} e^{-2\pi\Gamma_S(\tau - (\tau_0/2))} \sin c(i\pi\tau_0 \Gamma_S) & \tau \geq \tau_0/2 \\ e^{2\pi\Gamma_I(\tau - (\tau_0/2))} \sin c(i\pi\tau_0 \Gamma_I) & \tau < \tau_0/2 \end{cases} \right|^2 \quad (1)$$

Where $E_s^{(\pm)}$ and $E_i^{(\pm)}$ are operators for signal and idler photons respectively, $\gamma_{S,I}$ are the cavity damping rates for signal (S) and idler (I), $\omega_{S,I}$ are the central frequencies of signal and idler photons respectively, $\tau_0$ is the propagation delay between signal and idler inside the conversion crystal, $\Gamma_{S,I} = \gamma_{S,I}/2 + im_{S,I}\Delta\omega_{S,I}$ with mode indices $m_{S,I}$ and free spectral range $\Delta\omega_{S,I}$. As the result of compensation, $\Delta\omega_S = \Delta\omega_I = \Delta\omega$ for our cavity.

The input power of the pump beam is 13.2 *mW*, the coupling coefficient to the cavity is about 60%, so the effective pump power is 7.92 *mW*. When we lock the cavity to horizontal polarization of the pump beam, the leaked power of the pump from the output mirror is 0.275 *mW*, the transmittance of the output mirror at 780 *nm* is 0.14%, so the circulating power inside the cavity is about 196 *mW*. We adjust the temperatures of the two crystals while keeping the cavity locking to the pump, and observe the single count of APD to find the resonance point of the down-converted field. The resonance single count of APD1 is 2600 per second, and the maximum coincidence in 300 seconds is 259, the average accidental coincidence is 98. We see a dramatically decreasing in single counts compared to the single-pass experiment. Two factors are responsible for it. The first one is due to a bigger beam waist (120 *μm*) inside the cavity. Another one is related to the spatial mode match of the down-converted photons to the cavity eigenmode, only partial of the down-converted photons can fulfill this condition.

The measured cross-correlation between signal and idler is showed in Figure 6. The delay is adjusted using a delay generator (DG535) with a step of 1 *ns*, each data point is accumulated for 300 seconds, the accidental coincidence is not subtracted in the figure. Equation (1) shows that the cross-correlation function of the signal and idler for multi-mode output should be an oscillation damping multi-peaks comb-like shape, the damping rate of the peak-values is the same as the cavity damping rate. The envelope of the peaks is damping as $e^{-2\pi\gamma_{S,I}|\tau|}$, we assume $\gamma_S = \gamma_I = \gamma$ for our cavity, so we can estimate the bandwidth of the cavity from the envelope of Figure 6. The decay time $T_C = 1.39/2\pi\gamma$, the full wide at half maximum (FWHM) measured from figure 4 is 27.7 *ns*, corresponding to the bandwidth $\gamma = 8MHz$ in our OPO. We can't get multi-peaks experimentally for the following reasons: the second APD is running in gated mode with a detection window of 2.5 *ns*, and the resolution time of our system is determined by the detection window of the second APD, and this time is larger than the roundtrip time (≈1 *ns*) of the photons inside the cavity. The response of the whole detection system can be modeled using a Gaussian function

$$h(t) = \kappa e^{-\frac{t^2}{\sigma^2}} \quad (2)$$

Where $\kappa$ is a constant, and $\sigma$ characterize the effective detection window of the second APD (the value of $\sigma$ is determined in the single-pass experiment by adjusting the delay of the first APD and recording the counts of the second APD), so the measured coincidence count curve is the convolution of $G_{s,i}^{(2)}(\tau)$ and $h(t)$

$$\Gamma(t) = C \int_{-\infty}^{+\infty} G_{s,i}^{(2)}(\tau) h(t-\tau) d\tau \quad (3)$$

Where C is a constant. The black solid curve in Figure 6 is the numerical simulation of Equation (3) with parameters $\gamma_S = \gamma_I = \gamma = 8MHz$, $\Delta\omega = 0.952GHz$, $\sigma = 1.4ns$. the sums in Eq. (1) are evaluated for 255 longitude modes. The FWHM of the simulation curve is 27.7 ns, which agrees very well with the measured results of our experiment as shown in Figure 6.

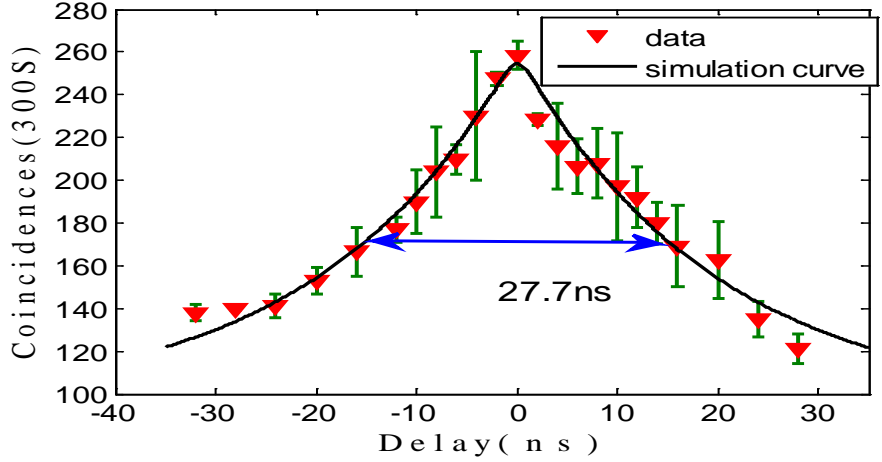

**Figure 6.** Coincidence count per 300 seconds as the function of the signal-idler delay. Red triangle points are experimental data-sets, black solid line is the simulation curve use Eq. (3) with parameters of our experiment.

The influence of the pump power to our source is showed in Figure 7. The relative delay is adjusted at the position of maximum coincidences. Figure 7(a) is the coincidences per 200 seconds measured as a function of pump power. It shows that the coincidences are linearly proportional to the pump power and the accidental coincidences increase dramatically with the increase of pump power. Figure 5(b) is the signal to noise ratio (SNR), which is defined as the ratio of maximum coincidence to the minimum coincidence. The minimum coincidence is measured by adjusting the delay to be far away from the balanced position. As a function of power, the SNR increases as the pump power decreases. The reason is as following: the single count decreases as the pump power decreases, the accidental coincidences due to uncorrelated photons and multi photons decrease.

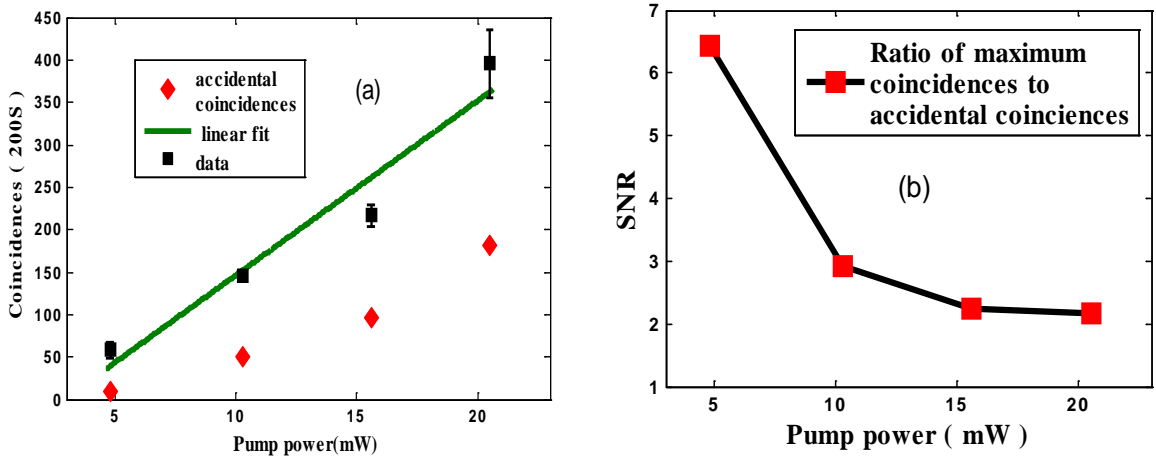

**Figure 7.** (a) coincidences per 200 seconds as the function of pump power. Black squares are measured data sets with accidental coincidences, red diamond data sets are corresponding accidental coincidences, green solid line is linear fit of data sets (b) SNR as the function of pump power.

The temperature detuning behavior of the output is depicted in Figure 8. This curve is obtained near a triple resonances point, we adjust the temperature of KTP crystal near this point while keep temperature of the PPKTP crystal unchanged and the pump resonant. The single counts per 10 seconds are measured as we adjust the temperature with a step of 0.1K. From the Lorentz fit to the data sets, the FWHM of the temperature dependence is 0.2K. Shifting from one longitude mode to the next mode corresponding to a cavity optical length change of $\lambda/2$, where $\lambda=1560nm$ is the wavelength of the down-converted field, and the half value from the peak will lead to a change of $\lambda/(2F)$, where $F\approx 120$ is the finesse of the cavity. For our OPO, an optical length change of 6.5 *nm* of the cavity is required. Two processes compete with each other when we tune the temperature of the KTP crystal. As the cavity is locked to the pump beam, when the crystal temperature increases, the length of the cavity will decrease to maintain a constant optical length for pump field. Due to different thermal coefficients at different wavelengths, the optical length of the down-converted fields will change, and the down-converted fields will shift away from resonance. From the thermal coefficient of KTP crystal, we can calculate the FWHM of the temperature detuning. We see that the output damps very fast from the resonance when the temperature is detuning from the resonance point. We also observe other triple resonances points when we tune the temperature of either crystal by about 8℃, this is a result of cluster emission effect of the cavity SPDC. For details, one can refer to reference [26].

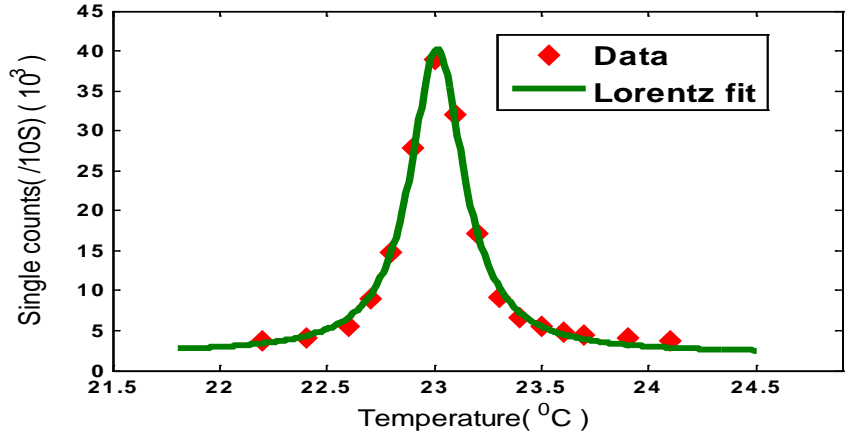

**Figure 8.** Single count per 10 seconds as the function of the temperature of the KTP crystal, the temperature of the PPKTP crystal keeps at a constant.

### 3.2 Single mode output

The single mode photon pairs are obtained using a filter cavity C2, the length of C2 is 5 mm (30 GHz FSR and finesse of 100) .We align the filter cavity using the diode laser, the same as the one used in Section 2, the wavelength of the laser is tuned to be exactly doubled respect to the pump wavelength. The cavity is set to be resonant by adjusting the temperatures of the two crystals and the frequency of the pump light to maximize the coincidences. The length of the cavity is stabilized by a temperature control unit. The input pump power is 20.6 *mW*, and the temporal cross-correlation of the signal and idler photon is measured ( Figure 9). The single count is about 890 per second, and the FWHM is about 27.7 *ns*, the solid line is the fitting using the function $e^{-2\pi\gamma|\tau|}$ with $\gamma=8MHz$. The using of a filter cavity should give single mode output of the source in principle, but due to relatively low finesse of the filter cavity, other mode may not be removed completely, and mix in the final output. As the number of longitude modes inside the cavity is large (about 255 estimated in our experiment), the filtered output photon number will be very small for a perfect filter cavity. The single count reduces to a ratio of 0.15 relative to the multi-mode output in our experiment due to imperfect cavity filter.

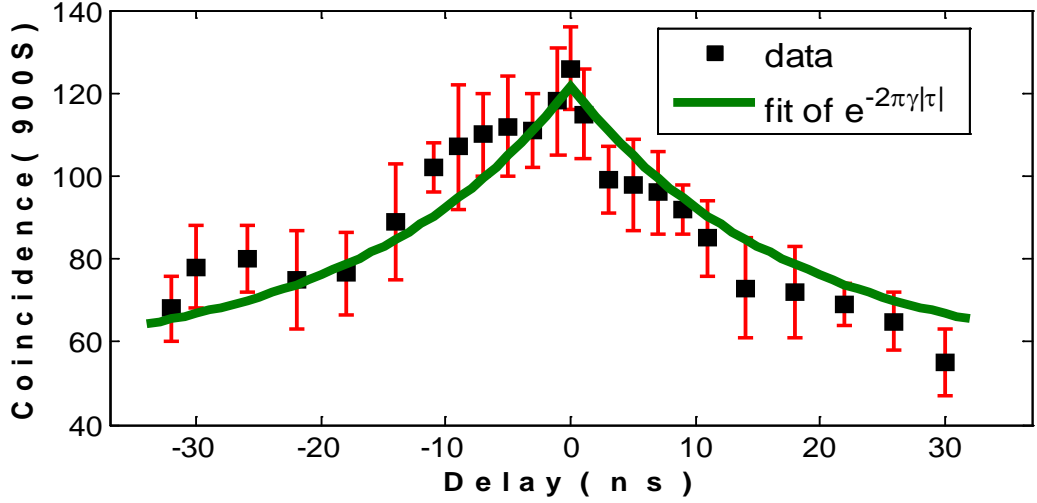

**Figure 8.** Coincidence counts with accidental coincidences per 900 seconds as the function of signal-idler delay. The solid line is fitted using function $e^{-2\pi\gamma|\tau|}$.

### 3.3 Estimation of the pair generation rate

The losses of each element in our system is characterized using a diode laser (DL prodesign, Toptica) running at 1560 *nm*. Table 1 lists the losses of each optical element. Accounts for all these loss factors, the estimates pair production rate is about 134 $s^{-1}MHz^{-1}mW^{-1}$ using formula

$$R_{estimate} = R_{detected} / (d\alpha^2 \alpha_1 \alpha_2 t_1^2 t_2^2 \eta^2) \text{, where } R_{detected} = 7.08\times 10^{-4}$$

(s · MHz · mW)$^{-1}$, $d = 2.5\%$ is the detection duty cycle.

**Table 1.** Value of different loss factors

| Loss factors | Value |
| --- | --- |
| Transmittance of long-pass filter $t_1$ | 0.78 |
| Transmittance of cavity C2 $t_2$ | 0.84 |
| Collecting efficiency of SMF $\alpha$ | 0.33 |
| Coupling efficiency of signal in FPBS $\alpha_1$ | 0.82 |
| Coupling efficiency of idler in FPBS $\alpha_2$ | 0.86 |
| Detection efficiency of APDs $\eta$ | 0.08 |

We compare the brightness of our source with main recently reported OPO sources in table 2. The high brightness of our source and the fiber-coupled single-mode emission, make it suitable for quantum communication applications. As the resonator is locked to the pump beam, the source can operate stably over several hours.

Table 2. Comparison between main recently reported OPO-based photon pair source according to different criteria

|  | Wang[16] | Kuklewicz[17] | Bao[21] | Scholz[24] | Wang[22] | Wolfgramm[25] | Pomarico[26] | our source |
|---|---|---|---|---|---|---|---|---|
| Wavelength (nm) | 860 | 795 | 780 | 893.4 | 780 | 795 | 1560 | 1560 |
| Bandwidth(MHz) | 18 | 22 | 9.6 | 2.7 | 20 | 7 | 117 | 8 |
| Single mode output |  | ✓ | ✓ | ✓ |  |  | ✓ | ✓ |
| Coupling into fiber |  | ✓ | ✓ | ✓ | ✓ | ✓ | ✓ | ✓ |
| Entanglement generation | ✓ | ✓ | ✓ |  |  | ✓ | ✓ |  |
| Spectral brightness* $((s \cdot MHz \cdot mW)^{-1})$ | 0.12 | 0.7 | 6 | 330 | 5.4 | 70 | 17 | 134 |

* All the brightness listed here are inferred.

## 4. Conclusions

A cavity-enhanced narrowband single-mode photon source at telecom wavelength is prepared with a type-II PPKTP crystal. Our source has stability of hours as the cavity is locked to the pump beam. A high spectral brightness of 134(s •MHz •mW)$^{-1}$ is achieved due to resonance of the pump beam. The bandwidth of the photon is 8 MHz, and the coherence time measured is 27.7 ns. The collected number of photons is highly affected by losses of external optical elements. By properly designing the photon collecting zoom-lens and using low loss filter elements like high reflectivity narrow bandwidth fiber grating filters, the loss would reduce drastically. An increase of 10 dB of collected photon pairs will be achieved for fiber coupling efficiency of 0.8 and filter transmittance of 0.95. Also the pump coupling efficiency can be improved further to reduce the pump power. Moreover, an update of our detection system with higher trigger rate (GHz) and higher efficiency (20%) single photon detectors will be of great benefit to reduce the data acquisition time. This source will be practical in quantum communication if we process the spectral of the photon using all fiber elements after collecting by single mode fiber with high efficiency.

## Acknowledgments

We thank Dr. Chen Wei for lending us single photon detector, and Prof. Xu Li-Xin for providing us with fiber elements. This work was supported by the National Natural Science Foundation of China (Grant Nos. 11174271, 61275115, 10874171), the National Fundamental Research Program of China (Grant No. 2011CB00200), and the Innovation fund from CAS, Program for NCET.